\documentstyle[twocolumn,aps,prb]{revtex}
\tighten
\begin{document}
\draft
\title{Few-anyon systems in a parabolic dot}
\author{Ricardo P\'erez$^1$\cite{ricardo} and 
 Augusto Gonzalez$^{1,2}$\cite{augusto}}
\address{$^1$Centro de Matem\'aticas y F\'\i sica 
 Te\'orica, Calle E No. 309, Ciudad Habana, Cuba\\
 $^2$Universidad Nacional, Sede Medell\'\i n,
 A.A. 3840, Medell\'\i n, Colombia}
\date{\today}
\maketitle

\begin{abstract}
The energy levels  of two and three anyons in a two-dimensional 
parabolic quantum dot and a perpendicular magnetic field are 
computed as power series in $1/|J|$, where $J$ is the angular 
momentum. The particles interact repulsively through a coulombic
($1/r$) potential. In the two-anyon problem, the reached accuracy 
is better than one part in $10^5$. For three anyons, we study the 
combined effects of anyon statistics and coulomb repulsion in the
``linear'' anyonic states.
\end{abstract}


\section{Introduction}

Recent developments in semiconductor technology (e.g. MBE 
and electron lithography) opened the possibility to create totally 
confined electron systems, the so called artificial atoms or quantum 
dots \cite{J95,JH97}. This is one of the examples of 
present-day-physics' interest in confined finite systems, among
which one can mention also the atomic and electronic traps\cite{traps},
and the condensation of confined bosons\cite{BE}.

Quantum dots exhibit very interesting properties like the possibility 
of varying their parameters (number of electrons, applied fields, 
dot's geometry, temperature) over a wide range, the observation of 
conductance oscillations\cite{K92} and Kohn's theorem\cite{MHP92},
etc.

Theoretically, these systems have been studied mainly with the help of
numerical methods. However, analytic approaches have proven to work 
extremely well providing, at the same time, a qualitative understanding
of the quantum dynamics. Among these approaches one can mention the 
semiclassical quantisation\cite{M96}, regularised perturbation 
theory\cite{MP94,GM97}, Pade approximant techniques\cite{G97,GPP97}, and
the $1/N$-expansion\cite{BLG89,QCG95,GPF97}.

In the present paper, we continue the analytic-qualitative line of 
research and apply the $1/N$-expansion ($N$ is the absolute value of 
the angular momentum) to compute the energy levels of two and three
anyons in a model parabolic dot. The particles interact through a 
coulombic ($1/r$) repulsive potential. A magnetic field is applied
perpendicularly to the plane of motion.

Numerical results for the two-anyon system were obtained by Myrheim 
et al\cite{MHV92}. Exact analytic solutions at particular values of 
the coupling constants were found in\cite{V91}. Bohr-Sommerfeld
quantisation was applied to this system at low magnetic fields\cite
{S97}. We shall see that our method provides extremely accurate 
solutions for states with angular momentum $|J|\ge 2$. A picture for 
the ``geometry'' of the states (the spatial distribution of probability)
is obtained also. In the three-anyon problem, however, to our knowledge
there are no numerical or approximate calculations. 

We start from the Hamiltonian of $N_a$ anyons moving in a two 
dimensional quantum dot in the presence of a perpendicular homogenous 
magnetic  field. In the bosonic gauge, it is given by the expression
\cite{M93}

\begin{eqnarray}
H &=& \frac 1{2m}\sum\limits_{i=1}^{N_a}\left| \vec{p_i}+\frac ec
\vec{A_i}-\hbar \nu \vec{a_i}\right| ^2+\frac m2\omega_0^2
\sum\limits_{i=1}^{N_a} r_i^2\nonumber\\
  &+& \sum\limits_{i>j}\frac{e^2}{\epsilon 
\left| \vec{r}_i-\vec{r}_j\right| },  
\end{eqnarray}

\noindent
in which the vector potential is taken in the symmetric gauge, 

\begin{equation}
\vec{A_i}=\frac 12\vec{B}\times \vec{r_i},
\end{equation}

\noindent
$\vec{a_i}$ is the statistical vector potential,

\begin{equation}
\vec{a_i}=\sum\limits_{j\neq i}\frac{\vec{n}\times
\left( \vec{r_j}-\vec{r_i}\right) }{\left| 
\vec{r_j}-\vec{r_i}\right| ^2},  
\end{equation}

\noindent
$\vec{n}$ is the unit vector perpendicular to the plane of
motion of the anyons, $e$ is the anyon's charge,
$\nu $ is the anyonic parameter, $\omega _0$ 
is the frequency of the parabolic potential needed to confine the 
anyons in the dot and $\epsilon $ is the dielectric constant 
of the medium. A dimensionless Hamiltonian is obtained by means 
of the change of variables $\vec{r_i}\rightarrow \sqrt{\hbar/ 
(m\Omega)}~\vec r_i$

\begin{eqnarray}
\frac H{\hbar \Omega } &=&\frac{1}{2}\sum\limits_{i=1}^{N_a} p_i^2+
\frac{\omega _c}{2\Omega }\vec{n}\cdot \sum\limits_{i=1}^{N_a} 
\vec{r_i}\times \vec{p_i}\nonumber\\
&+& \nu \vec{n}\cdot \sum\limits_{i>j}\frac
{\left( \vec{r_i}-\vec{r_j}\right) \times \left( \vec{p_i}-
\vec{p_j}\right) }{\left| \vec{r_i}-\vec{r_j
}\right| ^2}\nonumber\\
&+& \frac 12 \sum\limits_{i=1}^{N_a} r_i^2+
\frac{\omega_c}{4\Omega}\nu N_a(N_a-1) \nonumber\\
&+&\frac{\nu ^2}{2}\sum\limits_{i\neq j,k }\frac{\left( 
\vec{r_i}-\vec{r_j}\right) \cdot \left( 
\vec{r_i}-\vec{r_k}\right) }
{\left| \vec{r_i}-\vec{r_j}\right| ^2 
\left| \vec{r_i}-\vec{r_k}\right| ^2}
+\beta ^3 \sum\limits_{i>j}\frac{1}{\left| \vec{r_i}-\vec{r_j}
\right| }, 
\end{eqnarray}

\noindent
where $\omega _c=eB/(mc)$ is the cyclotronic frequency, 
$\Omega ^2=\omega _c^2/4+\omega _0^2$, and $\beta^3=
\sqrt{me^4/(\epsilon ^2 \Omega \hbar ^3)}$ is the square root of 
the ratio between the coulombic and oscillator characteristic 
energies. The problem has one exactly solvable limit: a low-density 
limit, which we call the Wigner limit, reached when $\beta \to \infty$.
In the $\beta \to 0$ (oscillator) limit, the two-anyon problem is 
exactly solvable \cite{2anyons}, whereas the three-anyon system has
an infinite family of exact linear states \cite{3anyons}. In real 
semiconductors, $\beta \sim 1$. 

Introducing Jacobi coordinates,
\begin{eqnarray}
\vec{\rho}_k &=&\sqrt{\frac{2k}{k+1}} \left \{ \frac{1}{k}
\sum\limits_{i=1}^k \vec{r_i}-\vec r_{k+1} \right \},
\nonumber\\
&&~~~1 \leq k \leq N_a-1, \\
\vec{\rho}_{N_a} &=&\frac{1}{\sqrt{N_a}}\sum\limits_{i=1}^{N_a} 
\vec{r_i},
\end{eqnarray}

\noindent
the centre-of-mass and relative motions are separated

\begin{eqnarray}
\frac H{\hbar \Omega }=\frac{H_{CM}}{\hbar \Omega }+
\frac{H_{rel}}{\hbar \Omega },
\end{eqnarray}

\noindent
where

\begin{eqnarray}
\frac{H_{CM}}{\hbar \Omega }&=&\frac{1}{2}p_{N_a}^2+
\frac{\omega_c}{2\Omega}\vec n \cdot 
\left(\vec {\rho}_{N_a} \times \vec p_{N_a} \right)+\frac{1}{2} 
\rho_{N_a}^2,
\end{eqnarray}

\noindent
is the centre of mass Hamiltonian and

\begin{eqnarray}
\frac{H_{rel}}{\hbar \Omega }&=&\sum\limits_{i=1}^{N_a-1} 
p_i^2+\frac{\omega _c}{2\Omega }\vec{n}\cdot \sum\limits_{i=1}^{N_a-1} 
\vec{\rho_i}\times \vec{p_i}\nonumber\\
&+&\nu \vec{n}\cdot \sum\limits_{i>j}
\frac{ \vec r_{ij} \times  \vec p_{ij}}{r_{ij}^2}+\frac 14 
\sum\limits_{i=1}^{N_a-1} \rho_i^2+\frac{\omega_c}{4\Omega}
\nu N_a(N_a-1)\nonumber \\
&+&\frac{\nu ^2}{2}\sum\limits_{i\neq j,k }\frac{\vec r_{ij} \cdot 
\vec r_{ik}}{r_{ij}^2 r_{ik}^2}+
\beta ^3 \sum\limits_{i>j}\frac{1}{r_{ij}}, 
\end{eqnarray}

\noindent
is the Hamiltonian of the relative motion. We introduced the
following notation: $\vec{r_{ij}}=\vec{r_i}-\vec{r_j}$, and 
$\vec{p_{ij}}=\vec{p_i}-\vec{p_j}$.

We will obtain approximate expressions for the energy eigenvalues of
$H_{rel}/(\hbar \Omega)$ for $N_a=2$ and $N_a=3$ as a function of 
$\beta$ by means of the $1/|J|$-expansion. 

\section{The two-anyon system}

In the two-anyon problem, we have only one Jacobi coordinate,
$\vec \rho_1$, and the Hamiltonian of the internal motion reads

\begin{eqnarray}
\frac {H_{rel}}{\hbar \Omega } &=& p_1^2+\frac{\omega _c}{2\Omega }
\vec n\cdot \left ( \vec \rho _1\times \vec p_1 \right )+2\nu \vec n
\cdot\frac{\vec \rho _1\times \vec p_1}{\rho _1^2}+\frac 14 
\rho _1^2\nonumber\\
&+& \frac {\nu ^2}{\rho _1^2}+\frac {\beta ^3} {\rho _1} +
\frac{\omega _c\nu }{2\Omega }.  
\end{eqnarray}

Notice that $\vec n\cdot\left(\vec\rho_1\times \vec p_1 \right)=J$.
After the scaling transformation $\rho _1^2\rightarrow \left| 
J\right| $ $\rho _1^2$, we get 

\begin{eqnarray}
h&=&\frac 1{\left| J\right| }\left[ \frac {H_{rel}}{\hbar \Omega }-
\frac{\omega _c (J+\nu )}{2\Omega }\right] =
\frac {(\tilde\nu+1)^2}{\rho _1^2}+\frac 14 \rho _1^2 +
\frac {\tilde \beta ^3}{\rho _1}\nonumber\\
&-& \frac 1{J^2} \left( \frac{\partial ^2}{\partial \rho _1^2}+
\frac 1{\rho _1}\frac \partial {\partial \rho _1} \right) ,  
\end{eqnarray}

\noindent
where $\rho _1=\left| \vec \rho _1\right|$. We have ``renormalised"
the coupling constant, $\tilde \beta ^3=\beta ^3/\left| J\right| ^{
\frac 32}$, and the statistical parameter, $\tilde\nu=\nu/J$, in order 
to take account of the Coulomb repulsion and the statistical 
interaction in a nonperturbative way when taking the formal limit 
$\left| J\right| \rightarrow \infty $. We shall look for symmetric 
eigenfunctions of $h$, i.e. $|J|$ shall be even.

In the $|J|\to\infty$ limit, the only term surviving in the 
Hamiltonian is the effective (classical) potential energy

\begin{equation}
U_{eff}=\frac{(\tilde\nu+1)^2}{\rho_1^2}+\frac{1}{4}\rho_1^2+
  \frac{\tilde\beta^3}{\rho_1}.
\end{equation}

Minimising $U_{eff}$, we obtain the leading contribution to the 
energy, $\epsilon_0=U_{eff}(\rho_{01})$, where the radius of the 
``Bohr orbit'' is obtained from

\begin{eqnarray}
\frac 12 \rho_{01}^4-\tilde\beta^3 \rho_{01}=2 (\tilde\nu+1)^2.
\end{eqnarray}

Substituting $\rho_1=\rho_{01}+y_1/|J|^{1/2}$ in the r.h.s. of (11)
and expanding, we get

\begin{equation}
h=\sum \limits_{i=0}^{\infty} \frac{h_i}{|J|^{i/2}},
\end{equation}

\noindent
where the operator coefficients are given by

\begin{eqnarray}
h_0 &=&\frac 34\rho _{01}^2-\frac {(\tilde\nu+1)^2}{\rho _{01}^2}, \\
h_1 &=&0, \\
h_2 &=&-\frac{\partial ^2}{\partial y_1^2}+\frac 14\left( 3+
\frac {4(\tilde\nu+1)^2}{\rho _{01}^4} \right)y_1^2 \\
h_i&=&(-1)^i \left \{ \left (\frac 1{2\rho_{01}^{i-2}}+
\frac {(i-1)(\tilde\nu+1)^2}{\rho_{01}^{i+2}} \right)y_1^i
\right.\nonumber\\
&+& \left.
\frac 1{\rho_{01}^{i-2}}y_1^{i-3}\frac{\partial}{\partial y_1} 
\right \},~~~i \geq 3.
\end{eqnarray}

Similar series are written for the wave function and the
scaled energy, that is, 

\begin{eqnarray}
\psi &=& \sum \limits_{i=0}^{\infty} \frac{\psi_i}{|J|^{i/2}},\\
\epsilon &=& \frac{1}{|J|} \left [\frac{E_{rel}}{\hbar \Omega}-
\frac{\omega_c (J+\nu)}{2 \Omega} \right ]=\sum \limits_{i=0}^
{\infty} \frac{\epsilon_i}{|J|^{i/2}}.
\end{eqnarray}

Inserting (19), (20) and the series expansion for the
Hamiltonian into the Sch\"odinger equation, we may compute the 
coefficients $\psi_i$ and $\epsilon_i$ in a systematic way. Up to 
second order, for example, the system is described in terms
of small oscillations around the equilibrium orbit, i.e. the 
wave function is

\begin{equation}
\Psi_0=e^{\imath J\theta}\mid n \rangle,
\end{equation}

\noindent
where $\theta$ is the angle associated to the vector $\vec\rho_1$,
the $\mid n \rangle$ are two-dimensional harmonic oscillator
radial states with frequency 
$\omega _1=\sqrt{3+4(\tilde\nu+1)^2/\rho _{01}^4}$,
and the first two coefficients for the energy are

\begin{eqnarray}
\epsilon _0 &=&\frac 34\rho _{01}^2-\frac {(\tilde\nu+1)^2}
{\rho _{01}^2}, \\
\epsilon _2 &=&\omega _1\left( n+\frac 12\right).
\end{eqnarray}

Afterward, we may take account of anharmonicities.  
The results for the next two coefficients are the following

\begin{eqnarray} 
  \epsilon _4 &=&-\frac 1{4 \rho_{01}^2}+ 
\frac {3 \left (2n^2+2n+1 \right)}{2\omega_1^2 \rho_{01}^2}
\left (1+\frac {6(\tilde\nu+1)^2}{\rho_{01}^4}\right)
\nonumber\\
&-& 
\frac {\left (30n^2+30n+11 \right)}{4\omega_1^4 \rho_{01}^2}
\left (1+\frac {4(\tilde\nu+1)^2}{\rho_{01}^4}\right)^2, \\
  \epsilon _6 &=&-\frac {3(2n+1)}{4 \omega_1 \rho_{01}^4}+
\frac {(2n+1)}{\omega_1^3 \rho_{01}^4}\left [5n^2+5n+9
\right. \nonumber\\
&+& \frac {(\tilde\nu+1)^2 \left (50n^2+50n+81\right )}
{\rho_{01}^4}\left.  \right]
-\frac {(2n+1)}{2 \omega_1^5 \rho_{01}^4}\left [87n^2
\right. \nonumber\\
&+& 87n+86+
\frac {12(\tilde\nu+1)^2 \left (87n^2+87n+86\right )}
{\rho_{01}^4}\nonumber\\
&+& \left.\frac {4(\tilde\nu+1)^4 \left (713n^2+713n+709\right )}
{\rho_{01}^8} \right]+ \frac {9(2n+1)}{2 \omega_1^7 \rho_{01}^4}
\nonumber\\
&\times& \left (25n^2+25n+19
\right ) \left (1+\frac {6(\tilde\nu+1)^2}{\rho_{01}^4}\right)
\nonumber\\
&\times& \left (1+\frac {4(\tilde\nu+1)^2}{\rho_{01}^4}\right)^2 
- \frac {15(2n+1)}{8 \omega_1^9 \rho_{01}^4}\left (47n^2+47n+31
\right ) \nonumber\\
&\times& \left (1+\frac {4(\tilde\nu+1)^2}{\rho_{01}^4}\right)^4.
\end{eqnarray}

Notice that in both Wigner ($\beta\to\infty$) and oscillator 
($\beta\to 0$) limits the corrections $\epsilon_4$ and $\epsilon_6$
go to zero. The expressions found in \cite{GPF97} are reproduced
if we take $\nu=0$.

We show in Fig. 1 the relative weight of $\epsilon_6$ in $\epsilon$ 
for the first states with $J=2$ and 6. The parameter $\nu$
was fixed to 1/2 (semions). The relative contribution of 
$\epsilon_6$ is never greater than $5\times 10^{-5}$
or $3\times 10^{-6}$ for $J=2$ or 6 respectively. This shows 
that the $1/|J|$-series is extremely well behaved.

A comparison with the exact solutions found in \cite{V91} is 
carried on in Fig. 2, where the relative difference $\left |
\epsilon-\epsilon_{exact}\right |/\epsilon_{exact}$ is 
plotted against $\nu$. The state with $J=6$, $n=0$ is shown. 
It may be easily verified that $\psi_{exact}=\rho_1^{|J+\nu|}
(1+\rho_1/\sqrt{2|J+\nu|+1})e^{-\rho_1^2}$, $\epsilon_{exact}=
|J+\nu|+2$, are exact solutions of the two-anyon problem at 
$\beta^3=\sqrt{2|J+\nu|+1}$. The comparison shows that the relative 
error of our estimate is not greater than $10^{-8}$.

\section{The Three-Anyon System}

The internal Hamiltonian of the system of three anyons in Jacobi 
coordinates $\vec {\rho}_1$ and $\vec{\rho}_2$ is written as

\begin{eqnarray}
  \frac {H_{rel}}{\hbar \Omega} &=&\sum\limits_{i=1}^2p_i^2+
\frac{\omega _c}{2\Omega}\vec n\cdot \sum\limits_{i=1}^2\vec 
\rho _i\times \vec p_i +\frac{3\omega _c\nu}{2\Omega} \nonumber\\
&+& \nu \vec n\cdot\left[ 2
\frac{\vec \rho _1\times \vec p_1}{\rho _1^2}+2\frac{\left( \vec 
\rho_1+\sqrt{3}\vec \rho _2\right) \times \left( \vec p_1+
\sqrt{3}\vec p_2\right) }{\left| \vec \rho _1+\sqrt{3}\vec 
\rho _2\right| ^2}\right.  \nonumber \\
&+& \left. 2\frac{\left( \vec \rho _1-\sqrt{3}\vec \rho _2\right) 
\times\left( \vec p_1-\sqrt{3}\vec p_2\right) }{\left| \vec 
\rho _1-\sqrt{3}\vec \rho _2\right| ^2}\right] +\frac 14
\sum\limits_{i=1}^2\rho _i^2 \nonumber\\
&+& 9\nu ^2\frac{\left( \rho _1^2+
\rho _2^2\right) ^2+4\left( \rho _1^2\rho _2^2-\left( \vec 
\rho _1\cdot \vec \rho _2\right) ^2\right) }{\rho _1^2\left| 
\vec \rho _1+\sqrt{3}\vec \rho _2\right| ^2\left| \vec \rho _1-
\sqrt{3}\vec \rho_2\right| ^2}+ \nonumber\\
&&\beta ^3\left[ \frac 1{\rho _1}+\frac 2{\left| \vec \rho _1+
\sqrt{3}\vec \rho _2\right| }+\frac 2{\left| \vec \rho _1-\sqrt{3}
\vec \rho _2\right|}\right].
\end{eqnarray}

Doing the same scaling transformation  
$\rho _i^2\rightarrow \left| J\right| \rho _i^2$, and making 
explicit the dependence on $|J|$, we get

\begin{eqnarray}
&& \frac 1{\left| J\right|} \left[ \frac {H_{rel}}{\hbar \Omega }-
\frac{\omega _c\left(J+3\nu \right) }{2\Omega }\right]  =
\frac 14\left( \frac 1{\rho _1^2}+\frac 1{\rho _2^2}\right) +
\frac 14\left( \rho _1^2+\rho _2^2\right) \nonumber\\
&+& \tilde \beta ^3\left[ 
\frac 1{\rho _1}+\frac 2{\left| \vec \rho _1+\sqrt{3}\vec \rho
_2\right| }+\frac 2{\left| \vec \rho _1-\sqrt{3}\vec \rho _2\right| }
\right] \nonumber \\
&+& 4\tilde\nu \frac{\rho _1^2\left( 2-3\cos ^2\theta \right) +3\rho
_2^2\left( 2-\cos ^2\theta \right) }{\left| \vec \rho _1+\sqrt{3}
\vec \rho_2\right| ^2\left| \vec \rho _1-\sqrt{3}\vec \rho _2
\right| ^2}+\frac {\tilde\nu}{\rho _1^2} \nonumber\\
&+& 9\tilde\nu ^2\frac{\left( 
\rho _1^2+\rho_2^2\right) ^2+4\rho _1^2\rho _2^2\sin ^2\theta }
{\rho _1^2\left| \vec \rho_1+\sqrt{3}\vec \rho _2\right| ^2\left| 
\vec \rho _1-\sqrt{3}\vec \rho_2\right| ^2} + 
\frac 1J \left[ \imath\left( \frac 1{\rho _1^2} 
- \frac 1{\rho _2^2}\right) \frac \partial {\partial \theta }
\right. \nonumber\\
&+& \tilde\nu \left( 12\imath \frac{\rho _1\rho _2\sin 2\theta }
{\left| \vec \rho _1+\sqrt{3}\vec \rho _2\right| ^2\left| 
\vec \rho _1-\sqrt{3}\vec \rho_2\right| ^2}\left( \rho _1
\frac \partial {\partial \rho _2}-\rho _2\frac \partial 
{\partial \rho _1}\right) \right.\nonumber\\
&+& \imath \frac 2{\rho _1^2}\frac \partial {\partial \theta }-
\left. \left. 8\imath \frac{\rho _1^2\left( 1-3\cos ^2\theta 
\right)+3\rho _2^2\left( 1+\cos ^2\theta \right) }{\left| 
\vec \rho _1+\sqrt{3}\vec \rho _2\right| ^2\left| \vec \rho _1-
\sqrt{3}\vec \rho _2\right| ^2}\frac \partial {\partial \theta }
\right)\right] \nonumber\\
&+& \frac 1{J^2}\left[ -\left( \frac{\partial ^2}
{\partial \rho _1^2}+\frac 1{\rho _1}
\frac \partial {\partial \rho _1}+\frac{\partial ^2}
{\partial \rho_2^2}+\frac 1{\rho _2}
\frac \partial {\partial \rho _2}+\right. \right. \nonumber\\
&&\left. \left. \left( \frac 1{\rho_1^2}+\frac 1{\rho _2^2}\right) 
\frac{\partial ^2}{\partial \theta ^2}\right)\right] , 
\end{eqnarray}

\noindent
where $\rho _1=\left| \vec \rho _1\right|$, $\rho _2=\left| \vec 
\rho_2\right|$, and $\cos \theta =\vec \rho _1
\cdot \vec \rho _2/(\rho_1\rho _2)$. The ``renormalised" 
$\tilde \beta ^3=\beta ^3/\left| J\right| ^{\frac 32}$ and 
$\tilde\nu=\nu/J$ were introduced.

The minimum of the classical potential entering (27) is reached in the 
configuration of an equilateral triangle ($\rho _{01}=\rho _{02}$,
$\theta =\pm \pi /2$). We choose, for example, $\rho _{01}=\rho _{02}$,
$\theta =\pi /2$. $\rho_{01}$ is obtained as the solution of 
the equation

\begin{equation}
\rho_{01}^4-3\rho_{01}\tilde\beta^3=(3\tilde\nu+1)^2.
\end{equation}

Then, introducing $\rho_1=\rho_{01}+y_1/\sqrt{|J|}$, 
$\rho_2=\rho_{01}+y_2/\sqrt{|J|}$ and $\theta=\pi /2 +z/\sqrt{|J|}$ 
in the r.h.s of (27), we obtain for the Hamiltonian $h$ a series 
like (14). The first operator coefficients are given by

\begin{eqnarray}
  h_0 &=&\frac 32\rho _{01}^2-\frac {(3\tilde\nu+1)^2}
{2\rho _{01}^2}, \\
  h_1 &=&0, \\
  h_2 &=&-\left( \frac{\partial ^2}{\partial y_1^2}+
\frac{\partial ^2}{\partial y_2^2}+\frac 2{\rho _{01}^2}
\frac{\partial ^2}{\partial z^2}\right) \nonumber\\
&-& \frac{2\imath }{\rho _{01}^3}{\rm sign}(J)\left( y_1-y_2\right) 
\frac \partial {\partial z}+\frac 14\left( 
\frac 3{\rho _{01}^4}+1\right) \left(y_1^2+y_2^2\right) \nonumber\\
&+& \frac 1{16}\left( 1-\frac 1{\rho _{01}^4}\right) \left(
5y_1^2+6y_1y_2+5y_2^2+3\rho _{01}^2z^2\right) \nonumber\\
&+& \frac{9\tilde\nu^2 }{16\rho_{01}^4}\left (y_1^2+y_2^2+6y_1y_2-
\rho_{01}^2z^2\right ) \nonumber \\
&+& \tilde\nu\left [\frac 3{8\rho_{01}^4}\left (
3y_1^2+3y_2^2+2y_1y_2 \right )-\frac {9z^2}{8\rho_{01}^2}
\right.\nonumber\\
&+& \frac {3\imath~{\rm sign}(J)z}{2\rho_{01}}\left (\frac{\partial}
{\partial y_1}-\frac{\partial}{\partial y_2}\right)\nonumber\\
&-& \left. 
\frac{3\imath~{\rm sign}(J)}{\rho_{01}^3}\left (y_1-y_2\right )
\frac{\partial}{\partial z}\right ], \\
  h_3 &=&-\frac 1{\rho _{01}}\left( \frac{\partial}{\partial y_1}+
\frac{\partial}{\partial y_2}\right) +\frac 2{\rho _{01}^3}\left(
y_1+y_2\right) \nonumber\\
&+& \frac{3\imath~{\rm sign}(J)}{\rho _{01}^4}\left(
y_1^2-y_2^2\right) \frac \partial {\partial z}-\frac 1{\rho _{01}^5}
\left(y_1^3+y_2^3\right) \nonumber\\
&-& \frac 1{64\rho _{01}}\left( 1-\frac 1{\rho _{01}^4}\right) \left(
19y_1^3+3y_1^2y_2+33y_1y_2^2+9y_2^2 \right. \nonumber\\
&-& \left. 9\rho _{01}^2y_1z^2+21\rho_{01}^2y_2z^2\right) 
- \frac {9\tilde \nu^2}{64 \rho_{01}^5}\left (y_2^3-5y_1^3
-21y_1^2 y_2 \right.\nonumber\\
&-& \left. 39 y_1 y_2^2+7\rho_{01}^2 y_1 z^2-11\rho_{01}^2 y_2 z^2 
\right) \nonumber\\
&+& \tilde\nu\left [\frac {3\imath~{\rm sign}(J)}{2 \rho_{01}^4} 
\left \{ \left (4y_1^2-2y_2^2-2y_1y_2-\rho_{01}^2z^2 \right) 
\frac {\partial}{\partial z} \right.\right.\nonumber\\
&-& \left. \rho_{01}^2 z\left (y_1 \frac{\partial}
{\partial y_2}+y_2 \frac{\partial}{\partial y_1} \right )+
2\rho_{01}^2 y_2 z \frac{\partial}{\partial y_2} \right\} 
\nonumber\\
&-& \frac{3}{32 \rho_{01}^5} \left (
21 y_1^3+15 y_2^3+5 y_1^2 y_2+23 y_1 y_2^2 \right.\nonumber\\
&+& \left.\left. \rho_{01}^2 y_1 z^2-13
\rho_{01}^2y_2 z^2\right )\right ],  \\
  h_4 &=&\frac 1{\rho _{01}^2}\left( y_1\frac{\partial}
{\partial y_1}+y_2\frac{\partial}{\partial y_2}\right) -
\frac 3{\rho _{01}^4}\left(y_1^2+y_2^2\right) \frac{\partial ^2}
{\partial z^2} \nonumber\\
&-& \frac{4\imath~{\rm sign}(J)}{\rho _{01}^5}\left( 
y_1^3-y_2^3\right) \frac \partial {\partial z}+\frac 5{4\rho _{01}^6}
\left( y_1^4+y_2^4\right) \nonumber\\
&+&\frac 1{256\rho _{01}^2}\left( 1-\frac 1{\rho _{01}^4}\right) 
\left( \frac{329}4y_1^4-25y_1^3y_2 \right.\nonumber\\
&+& \frac{123}2y_1^2y_2^2+135y_1y_2^3+
\frac 94y_2^4-\frac{99}2\rho _{01}^2y_1^2z^2 \nonumber \\
&+& 27\rho _{01}^2y_1y_2z^2+
\left. \frac{141}2\rho _{01}^2y_2^2z^2+\frac{41}4\rho _{01}^4z^4
\right) \nonumber\\
&+& \tilde\nu \left [-\frac {\imath~{\rm sign}(J)}{8 \rho_{01}^5}
\left\{ 6\left(11y_1^3-y_1^2y_2-7y_1y_2^2-3y_2^3
\right.\right.\right. \nonumber\\
&-& \left. 3 \rho_{01}^2y_1z^2-\rho_{01}^2
y_2z^2\right)\frac{\partial}{\partial z} \nonumber\\
&+& \rho_{01}^2 z\left(9y_1^2-3y_2^2-18y_1y_2-\rho_{01}^2z^2 
\right)\frac{\partial}{\partial y_1} \nonumber\\
&-& \left. \rho_{01}^2 z
\left(9y_1^2-27y_2^2+6y_1y_2-\rho_{01}^2z^2 \right)
\frac{\partial}{\partial y_2} \right\} \nonumber\\
&+& \frac{3}{512 \rho_{01}^6} \left (503y_1^4-28y_1^3y_2+
266y_1^2y_2^2+612y_1y_2^3 \right. \nonumber\\
&+& 183y_2^4-58 \rho_{01}^2y_1^2z^2+
20 \rho_{01}^2y_1y_2z^2-154 \rho_{01}^2y_2^2z^2 \nonumber\\
&-& \left.\left. 41 \rho_{01}^4z^4 \right) \right ] 
+ \frac{3\tilde\nu ^2}{1024\rho _{01}^6}\left (549y_1^4-411y_2^4-
84y_1^3y_2 \right.\nonumber\\
&+& 2718y_1^2y_2^2+1836y_1y_2^3-750\rho_{01}^2 y_1^2z^2+
1266\rho_{01}^2y_2^2z^2 \nonumber\\
&+& 60\rho_{01}^2y_1y_2z^2+ \left. 37\rho_{01}^4z^4\right ).  
\end{eqnarray}

The operator $h_2$ is diagonalised by changing variables 
$y_1=(y_s+y_m)/\sqrt{2}$, $y_2=(y_s-y_m)/\sqrt{2}$, 
$z=\sqrt{2} z_m/\rho_{01}$, and making the
``gauge'' transformation $h_2'=e^{\imath f}h_2e^{-\imath f}$ where
$f={\rm sign}(J)y_m z_m/(2 \rho_{01}^2)$, the results is

\begin{eqnarray}
h_2'=h_s+h_m,
\end{eqnarray}

\noindent
where $h_s$ describes the motion of a harmonic oscillator 
in the coordinate $y_s$ (the symmetric mode),

\begin{eqnarray}
h_s=-\frac{\partial ^2}{\partial y_s^2}+\frac{\omega_1^2}{4}y_s^2,
\end{eqnarray}

\noindent
with $\omega_1=\sqrt{3+(3\tilde\nu+1)^2/\rho _{01}^4}$,
and $h_m$ accounts for two dimensional motion in a ``fictitious''
magnetic field (the mixed mode)

\begin{eqnarray}
h_m &=& -\frac{\partial ^2}{\partial y_m^2}-\frac{\partial ^2}
{\partial z_m^2}+\frac{\omega_2^2}{4}(y_m^2+z_m^2) \nonumber\\
&-& \frac{\imath~{\rm sign}(J)(3\tilde \nu+1)}{\rho_{01}^2}\left (
y_m \frac{\partial}{\partial z_m}-z_m \frac{\partial}{\partial y_m} 
\right ),
\end{eqnarray}

\noindent
where $\omega _2=\sqrt{3/2-(3\tilde\nu+1)^2/(2\rho _{01}^4)}$.

The first two coefficients in the expansion for the energy are 

\begin{eqnarray}
\epsilon _0 &=&\frac 32\rho _{01}^2-\frac {(3\tilde\nu+1)^2}
{2\rho _{01}^2}, \\
\epsilon _2 &=&\omega _1\left( n_s+\frac 12\right) +
\omega _2\left(2n+\left| m\right| +1\right) +\omega _3~{\rm sign}(J)m,
\end{eqnarray}

\noindent
where $\omega_3=(3\tilde \nu+1)/\rho_{01}^2$, and the quantum 
numbers $n_s$, $n$, and $m$ may be used to approximately label 
the states.

Up to this order, the wave function is given by

\begin{equation}
\Psi_0=e^{i J\Xi} \mid J,n_s,n,m\rangle,
\end{equation}

\noindent
where $\Xi$ accounts for overall rotations of the system, and
$\mid J,n_s,n,m\rangle$ are the eigenfunctions of $h_2'$. We 
notice that, when $\beta\to 0$ the energy becomes

\begin{equation}
E_0=|J+3 \nu|+2+2 n_s+2 n+|m|+{\rm sign}(J) m.
\end{equation}

\noindent
These are the ``linear'' three-anyon states. We stress that they
are obtained as harmonic excitations against the equilateral
triangle configuration, and are not necessarily related to a 
cigar-like shape of the wave function ($\rho_1>>\rho_2$) as 
described in \cite{SVZ93}.

The set of numbers \{$J,n_s,n,m$\} compatible with the symmetry
constraints (the wave function shall be symmetric) are obtained
upon comparison with harmonic-oscillator wave functions at $\nu=0$,
$\beta=0$. Details may be found in \cite{GQR96,GPF97}. An additional 
requirement is that the state at $\beta=0$ should be a linear state.
For example, the lowest linear states are the following: $|0,0,0,0>$
(the g.s., starting from $E_0=2$ at the bosonic end), $|0,1,0,0>$,
and $|2,0,0,-1>$ (starting from $E_0=4$), $|3,0,0,0>$ and 
$|1,0,0,1>$ (starting from $E_0=5$), $|4,0,0,-2>$, $|2,1,0,-1>$, 
and $|0,1,1,0>$ (starting from $E_0=6$), etc. The lowest state with 
$J<0$ is $|-6,0,0,0>$, which starts at $E_0=8$. Of course, states
with small values of $|J|$ can not be described within our method.

In what follows, we restrict the analysis to levels with quantum
numbers $n_s=n=m=0$. This leaves only the linear anyonic states 
with $J=3 k$, where $k$ is an integer \cite{GQR96}. The geometry of 
the state is an equilateral triangle. It can be seen from (29) that
the side of the triangle increases with $\nu$ when $J>0$, and 
decreases when $J<0$. Thus, the coulomb repulsion is much more 
stronger for $J<0$ states, and the ordering of levels may 
dramatically change as $\beta$ is increased. On the other hand, 
for $\beta\to\infty$ the  side grows like $\rho_{01}\sim 3^{1/3}
\tilde \beta$ and becomes independent of the anyonic parameter, as
one expects. A strong coupling expansion \cite{G97} shows that the 
leading contribution to the energy (potential energy) is $\sim 
\beta^2$, the next corrections (quantum fluctuations) are $\sim 1$, 
the angular momentum and the statistical parameter enter the 
second order corrections, which are $\sim 1/\beta^2$.

The first anharmonic corrections to the energy are given by

\begin{eqnarray}
\epsilon_4 &=&\langle J,n_s,n,m\mid h_4^{\prime}\mid J,n_s,n,m
\rangle \nonumber\\
&+& \sum\limits_{n_s^{\prime },n^{\prime },m^{\prime }}
\frac{\langle J,n_s,n,m\mid h_3^{\prime }\mid J,n_s^{\prime},
n^{\prime },m^{\prime}\rangle }
{\epsilon _2^{Jn_snm}-\epsilon _2^{Jn_s^{\prime}n^{\prime}
m^{\prime }}} \nonumber\\
&\times& \langle J,n_s^{\prime },n^{\prime },
m^{\prime }\mid h_3^{\prime}\mid J,n_s,n,m\rangle,
\end{eqnarray}

\noindent
where $h'_3$ and $h'_4$ are obtained from $h_3$ and $h_4$ by means
of a gauge transformation, in the same way as explained above for 
$h_2$.
  
For a state with quantum numbers $\mid J,0,0,0\rangle$, we get 

\begin{eqnarray}
&&\langle J,0,0,0 \mid h_4^{\prime }\mid J,0,0,0\rangle =
\nonumber\\
&-& \frac 5{8\rho_{01}^2}+\frac 34\frac 1{\rho _{01}^2\omega _1^2}+
\frac 98\frac 1{\rho_{01}^6\omega _1^2}+\frac 9{16}
\frac 1{\rho _{01}^2 \omega _1\omega _2}
\left(1+\frac 1{\rho _{01}^4}\right)  \nonumber\\
&+& \frac 38 \frac{\omega _2}{\rho _{01}^2\omega_1}+ 
\frac 9{16}\frac 1{\rho _{01}^2\omega _2^2}\left( 1-\frac 1
{\rho _{01}^4}\right) \nonumber\\
&+& \frac{27\tilde\nu ^2}{16\rho _{01}^6 \omega_1^2 
\omega_2^2} \left ( 3\omega_1^2-\omega_1\omega_2+6\omega_2^2 \right)
\nonumber\\
&-& \frac{3\tilde\nu}{8\rho _{01}^6 \omega_1^2 \omega_2^2} \left(
9\omega_1^2-\omega_1\omega_2-18\omega_2^2 \right), \\
h_3^{\prime }&=& A\frac \partial {\partial y_s}+By_s+Cy_s^3+D,
\end{eqnarray}

\noindent
where 

\begin{eqnarray}
  A &=&-\frac{\sqrt{2}}{\rho _{01}}-\frac{3\sqrt{2} \imath~{\rm sign}(J) 
\tilde\nu}{4 \rho_{01}^3}\xi^2 \sin 2\alpha, \\
  B &=&\frac{\sqrt{2}}{\rho _{01}}\left\{ \sin ^2\alpha 
\frac{\partial ^2}{\partial \xi ^2}+\frac{\cos ^2\alpha }{\xi ^2}
\frac{\partial ^2}{\partial\alpha ^2} \right. \nonumber\\
&+& \left( \frac{(3\tilde\nu+2)
\imath~{\rm sign}(J)}{\rho _{01}^2}\cos ^2\alpha -\frac{\sin 2\alpha }
{\xi ^2} \right. \nonumber\\
&+& \left. \frac{3\tilde\nu\imath~{\rm sign}(J)}{2\rho _{01}^2}\right) 
\frac \partial {\partial \alpha } \nonumber\\
&+& \left( \frac{(3\tilde\nu+2)\imath~{\rm sign}(J)\sin 2\alpha 
\xi }{2\rho _{01}^2}+\frac{\cos ^2\alpha }\xi
\right) \frac \partial {\partial \xi } \nonumber\\
&+& \frac{\sin 2\alpha }\xi 
\frac{\partial ^2}{\partial \xi \partial\alpha }+
\frac {9\tilde\nu^2-1}{4\rho _{01}^4}\xi ^2\cos ^2\alpha \nonumber\\
&-& \left. \frac 3{16}\left( 1+\frac {3\tilde\nu^2-2\tilde\nu-1}
{\rho _{01}^4}\right) \xi ^2 \right\} , \\
  C &=&-\frac{\sqrt{2}}{4\rho _{01}}\left( 1+\frac {(3\tilde\nu+1)^2}
{\rho _{01}^4}\right) , \\
  D &=&-\frac{\sqrt{2}}{32\rho _{01}}\left( 5+
\frac {27\tilde\nu^2-6\tilde\nu-5}{\rho _{01}^4}\right)
\xi ^3\cos \alpha \left( 4\cos ^2\alpha -3\right)\nonumber\\
&+& \frac{3 \sqrt{2} \imath \tilde\nu~{\rm sign}(J)}
{2 \rho_{01}^3}\xi^2 \sin \alpha (4\cos^2 \alpha-1) 
\frac{\partial}{\partial \xi} \nonumber\\
&+& \frac{3 \sqrt{2} \imath \tilde\nu~{\rm sign}(J)}
{2 \rho_{01}^3}\xi \cos \alpha (4\cos^2 \alpha-3) \frac{\partial}
{\partial \alpha}.
\end{eqnarray}

Polar coordinates have been introduced according to 
$\xi^2=y_m^2+z_m^2$, $\tan \alpha=z_m/y_m$. The only nonvanishing
matrix elements entering the sum (41) are the following
 
\begin{eqnarray}
  \langle 0,0 \mid A\mid 0,0\rangle &=& -\frac{\sqrt{2}}{\rho_{01}},\\
  \langle 0,0 \mid A\mid 0,\pm 2\rangle &=& \pm 
\frac{3~{\rm sign}(J) \tilde\nu}{2\rho_{01}^3 \omega_2},\\
  \langle 0,\pm 2 \mid A\mid 0,0\rangle &=& \mp 
\frac{3~{\rm sign}(J) \tilde\nu}{2\rho_{01}^3 \omega_2},\\
  \langle 0,0 \mid B\mid 0,0\rangle &=&
-\frac{\sqrt{2}\omega _2}{2\rho _{01}}, \\
  \langle 0,\pm 2 \mid B\mid 0,0\rangle &=& \mp~{\rm sign}(J)
\frac {3\tilde\nu+2}{2\rho _{01}^3}-
\frac 14 \frac{\omega _2}{\rho _{01}} \nonumber\\
&+& \frac {9\tilde\nu^2-1}{4} 
\frac {1}{\omega _2\rho _{01}^5},\\
  \langle 0,\mp 3 \mid D\mid 0,0\rangle &=& -
\frac{\sqrt{6}}{16\rho _{01}^5\omega
_2^{3/2}}\left( 5 \rho_{01}^4-5 \right. \nonumber\\
&\mp& \left. 24 \rho_{01}^2~{\rm sign}(J) \tilde
\nu\omega_2+27\tilde\nu^2-6\tilde\nu \right) .
\end{eqnarray}

Collecting everything, we arrive to

\begin{eqnarray}
\epsilon _4 &=&\langle J,0,0,0\mid h_4^{\prime }\mid J,0,0,0
\rangle-\left \{ \frac{\langle 0,3\mid D\mid 0,0\rangle^2}
{3\omega_2+3\omega_3~{\rm sign}(J)}\right. \nonumber\\
&+& \left. \frac{\langle 0,-3\mid D\mid 0,0\rangle^2}
{3\omega_2-3\omega_3~{\rm sign}(J)}
\right \}-\frac{11}{\omega_1^4}
\langle 0,0\mid C\mid 0,0\rangle^2 \nonumber\\  
&+&\frac{\omega_1}{4}\left \{
\frac{\langle 0,0\mid A\mid 0,0\rangle^2}{\omega_1}+ 
\frac{\langle 0,0\mid A\mid 0,2\rangle 
\langle 0,2\mid A\mid 0,0\rangle}
{\omega_1+2\omega_2+2\omega_3~{\rm sign}(J)}\right.\nonumber\\
&+& \left. \frac{\langle 0,0\mid A\mid 0,-2\rangle 
\langle 0,-2\mid A\mid 0,0\rangle}
{\omega_1+2\omega_2-2\omega_3~{\rm sign}(J)} \right \} 
\nonumber \\
&-&\left \{ \frac{\langle 0,0\mid A\mid 0,2\rangle 
\langle 0,2\mid B\mid 0,0\rangle}
{\omega_1+2\omega_2+2\omega_3~{\rm sign}(J)}\right.\nonumber\\
&+& \left. \frac{\langle 0,0\mid A\mid 0,-2\rangle 
\langle 0,-2\mid B\mid 0,0\rangle}
{\omega_1+2\omega_2-2\omega_3~{\rm sign}(J)} \right \}\nonumber\\
&-& \frac{6}{\omega_1^3}\langle 0,0\mid B\mid 0,0\rangle
\langle 0,0\mid C\mid 0,0\rangle \nonumber \\
&-&\frac{1}{\omega_1}\left \{ 
\frac{\langle 0,0\mid B\mid 0,0\rangle^2}
{\omega_1}+\frac{\langle 0,2\mid B\mid 0,0\rangle^2}
{\omega_1+2\omega_2+2\omega_3~{\rm sign}(J)} \right.\nonumber\\
&+& \left. \frac{\langle 0,-2\mid B\mid 0,0\rangle^2}
{\omega_1+2\omega_2-2\omega_3~{\rm sign}(J)}
\right \}.
\end{eqnarray}

It may be checked that the corrections go to zero in both 
the Wigner $\left( \beta \rightarrow \infty \right)$
and the oscillator $\left( \beta \rightarrow 0\right)$ limits.

We show in Fig. 3 the relative weight of $\epsilon_4$ in
$\epsilon$ for three semions in states with $J=3$ and $J=6$. 
The numbers are similar to 
those appearing in the two-anyon problem. Thus, we expect a 
similar accuracy to this order, i.e. one part in $10^3$ or
better.

In Fig. 4, the levels with $J=\pm 6$ are drawn. $\beta$ is 
increased from 0.5 to 8. Notice that the coulomb effects are 
stronger for the state with negative $J$, and that the levels
become flatter (as a function of $\nu$) as $\beta$ rises.

In conclusion, the energy levels of two and three anyons in 
a model parabolic dot were computed by means of the $1/|J|$-
expansion. The qualitative picture emerging from the $1/|J|$-
expansion is that of a rigid structure (an orbit in the 
two-anyon system, an equilateral triangle for three anyons)
against which harmonic and anharmonic oscillations are 
developed. The coulomb repulsion is much stronger for negative-
$J$ states. Comparison with exact particular solutions for 
two anyons shows excellent agreement.
\vspace{1cm}

Acknowledgements: A. G. acknowledges support from the Colombian
Institute for Science and Technology (COLCIENCIAS).

\begin{figure}
\caption{Relative weight of $\epsilon_6$ in $\epsilon$.
 Two anyons in states with $n=0$ and $\nu=1/2$ are studied.
 a) $J=2$, b)$J=6$.}
\label{fig1}
\end{figure}

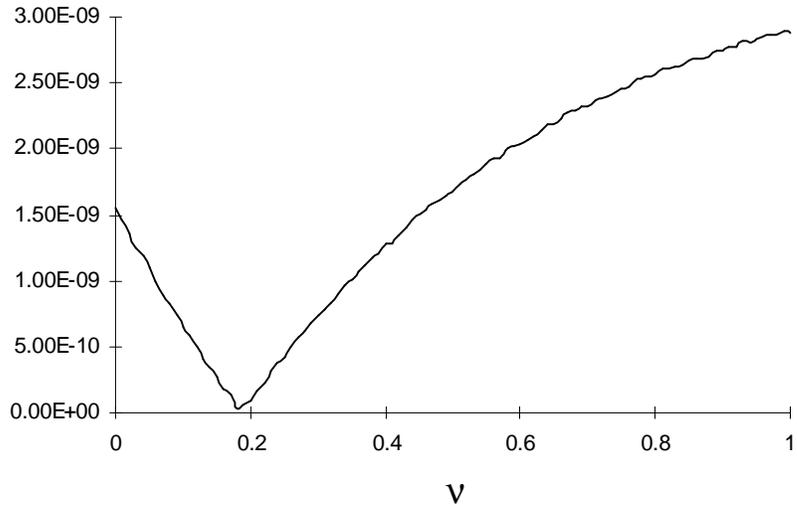
\begin{figure}
\caption{Comparison between the $1/|J|$-estimate and 
 the exact solution found in [16] for two anyons
 with $J=6$, $n=0$.}
\label{fig2}
\end{figure}

\begin{figure}
\caption{Relative weight of $\epsilon_4$ for three 
 anyons in states with $\nu=1/2$. a) $J=3$, b) $J=6$.}
\label{fig3}
\end{figure}

\begin{figure}
\caption{$|J| \epsilon$ vs $\nu$ for three anyons in 
 states with $J=\pm 6$. a) $\beta=0.5$, b) $\beta=8$.}
\label{fig4}
\end{figure}


\begin{thebibliography}{9}
\bibitem[\circ]{ricardo} Electronic mail: rperez@cidet.icmf.inf.cu
\bibitem[*]{augusto} Electronic mail: agonzale@perseus.unalmed.edu.co
\bibitem{J95} N. F. Johnson, J. Phys.: Condens. Matter
 {\bf 7}, 965 (1995).
\bibitem{JH97} J. H. Jefferson and W. H\"ausler, cond-mat/9705012.
\bibitem{traps} S. Becker et al., Rev. Sci. Instr. {\bf 66}, 
 4902 (1995); H. J. Kluge, Nucl. Instr. Methods Phys. Res. 
 {\bf B 98}, 500 (1995).
\bibitem{BE} C. C. Bradley, C. A. Sackett, J. J. Tollet and 
 R. G. Hulet, Phys. Rev. Lett. {\bf 75}, 1687 (1995); K. B. Davis, 
 M. O. Mewes, M. R. Andrews et al, Phys. Rev. Lett. {\bf 75}, 3969
 (1995).
\bibitem{K92} M. A. Kastner, Rev. Mod. Phys. {\bf 64},
 849 (1992).
\bibitem{MHP92} B. Meurer, D. Heitmann and K. Ploog, 
 Phys. Rev. Lett. {\bf 68}, 1371 (1992).
\bibitem{M96} P. A. Maksym, Phys. Rev. {\bf B 53}, 10871 (1996).
\bibitem{MP94} A. Matulis and F. M. Peeters, J. Phys.: Condens.
 Matter {\bf 6}, 7751 (1994).
\bibitem{GM97} A. Gonzalez, I. Mikhailov, cond-mat/9706013;
 Int. J. Mod. Phys. {\bf B 11} (1997), to appear.
\bibitem{G97} A. Gonzalez, J. Phys.: Condens. Matter {\bf 9}, 4643
 (1997).
\bibitem{GPP97} A. Gonzalez, B. Partoens, F. M. Peeters, 
 cond-mat/9707258; Phys. Rev. {\bf 56} (1997), to appear.
\bibitem{BLG89} A. Belov, Yu. Lozovik, A. Gonzalez, Phys. Lett. 
 {\bf A 142}, 389 (1989).
\bibitem{QCG95} L. Quiroga, A. Camacho, A. Gonzalez, 
 J. Phys: Condens. Matter {\bf 7}, 7517 (1995).
\bibitem{GPF97}  A. Gonzalez, R. P\'erez and P. Fileviez, 
 J. Phys.: Condens. Matter {\bf 9}, 8465 (1997).
\bibitem{MHV92} J. Myrheim, E. Halvorsen and A. Vercin, 
 Phys. Lett. {\bf B 278}, 171 (1992).
\bibitem{V91}  A. Vercin, Phys. Lett. {\bf B 260}, 120 (1991).
\bibitem{S97} L. Salasnich, Mod. Phys. Lett. {\bf B 11},
 269 (1997).
\bibitem{M93} J. Myrheim, Anyons, in Course on Geometric Phases, 
 ICTP, Trieste, 1993.
\bibitem{2anyons} J. M. Leinaas and J. Myrheim, Nuovo Cimento 
 {\bf 37 B}, 1 (1977).
\bibitem{3anyons} Y. S. Wu, Phys. Rev. Lett. {\bf 53}, 111 
 (1984); M. D. Jhonson and G. S. Canright, Phys. Rev. {\bf D 41},
 6870 (1990); A. P. Polychronakos, Phys. Lett. {\bf B 264}, 
 362 (1991).
\bibitem{SVZ93} M. Sporre, J. Verbaarschot and I. Zahed, Nucl. 
 Phys. {\bf B 389}, 645 (1993).
\bibitem{GQR96} A. Gonzalez, L. Quiroga and B. Rodriguez, 
 Few-Body Systems {\bf 21}, 47 (1996).
\end{thebibliography}
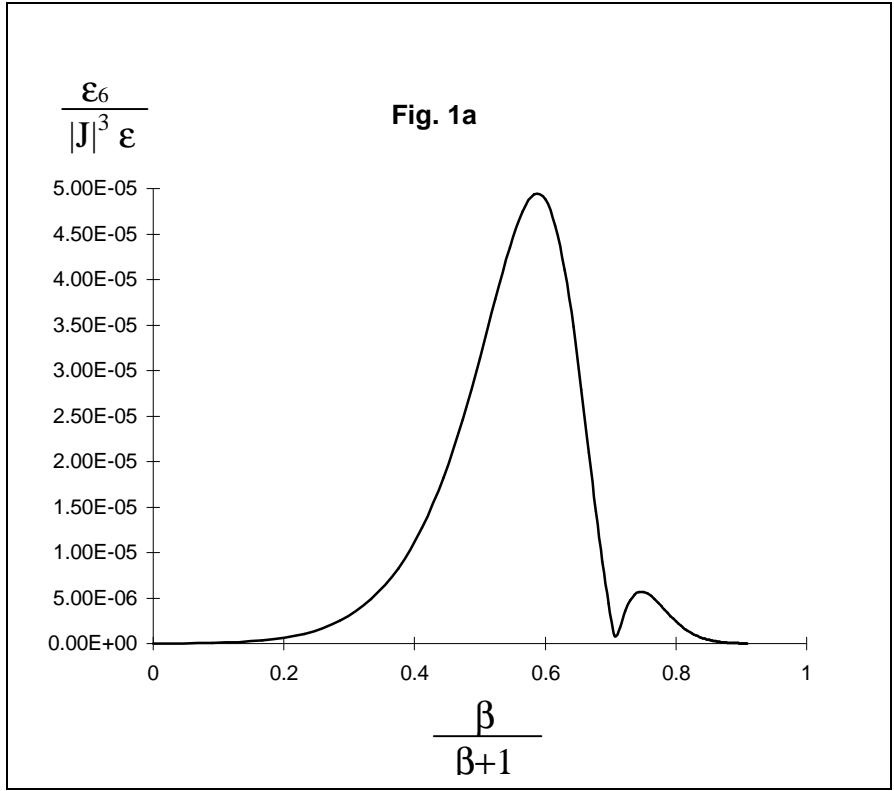
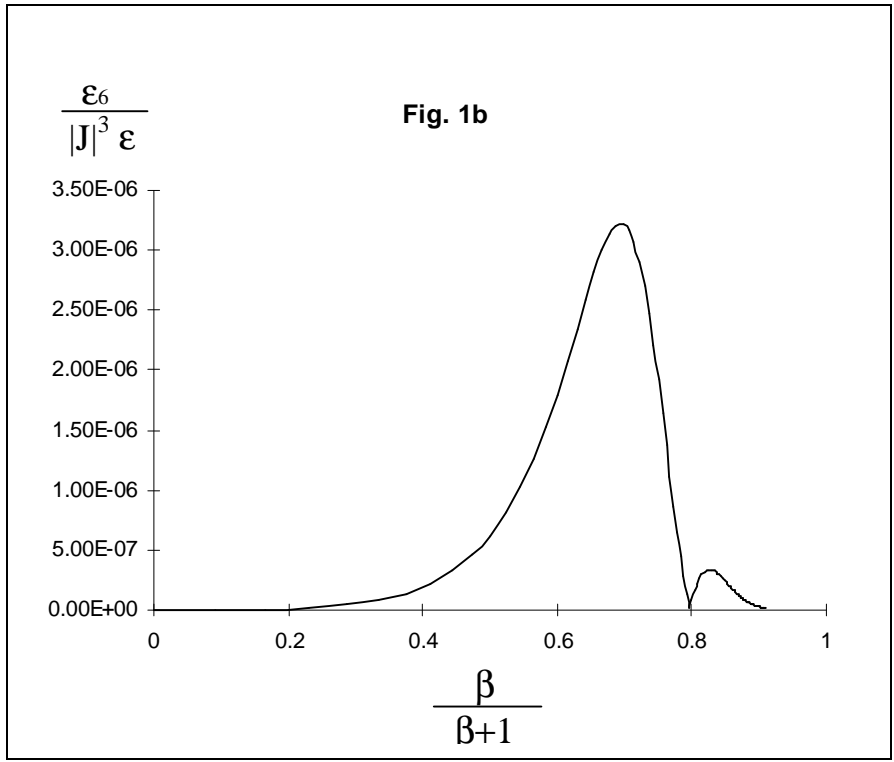
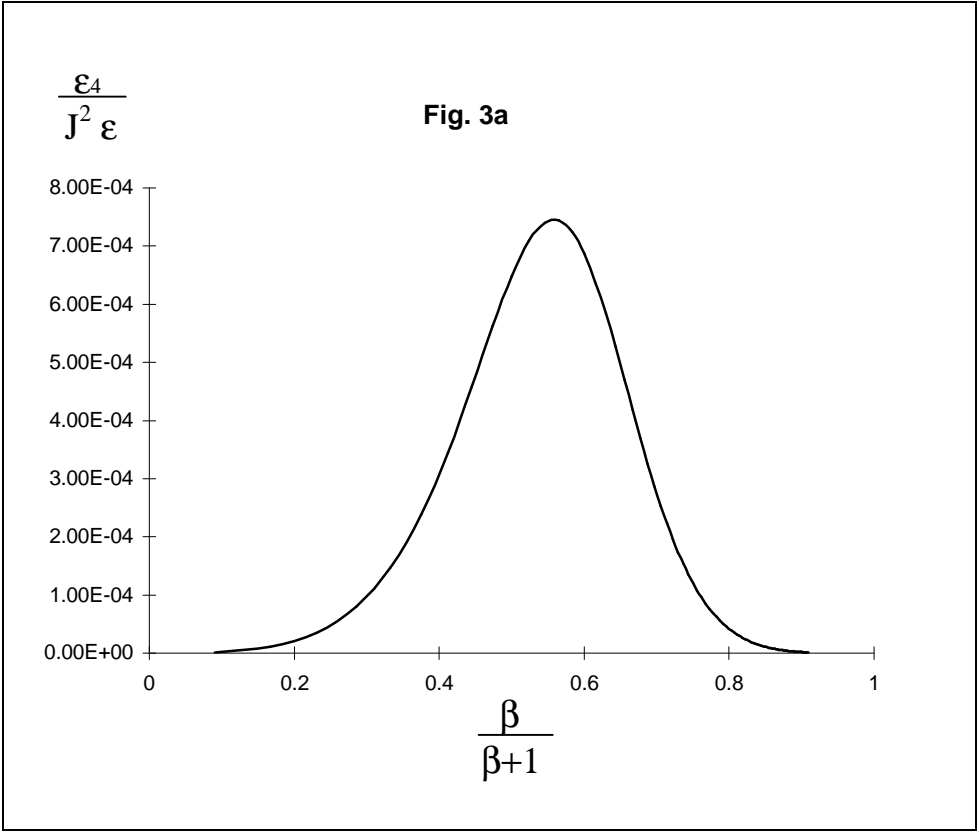
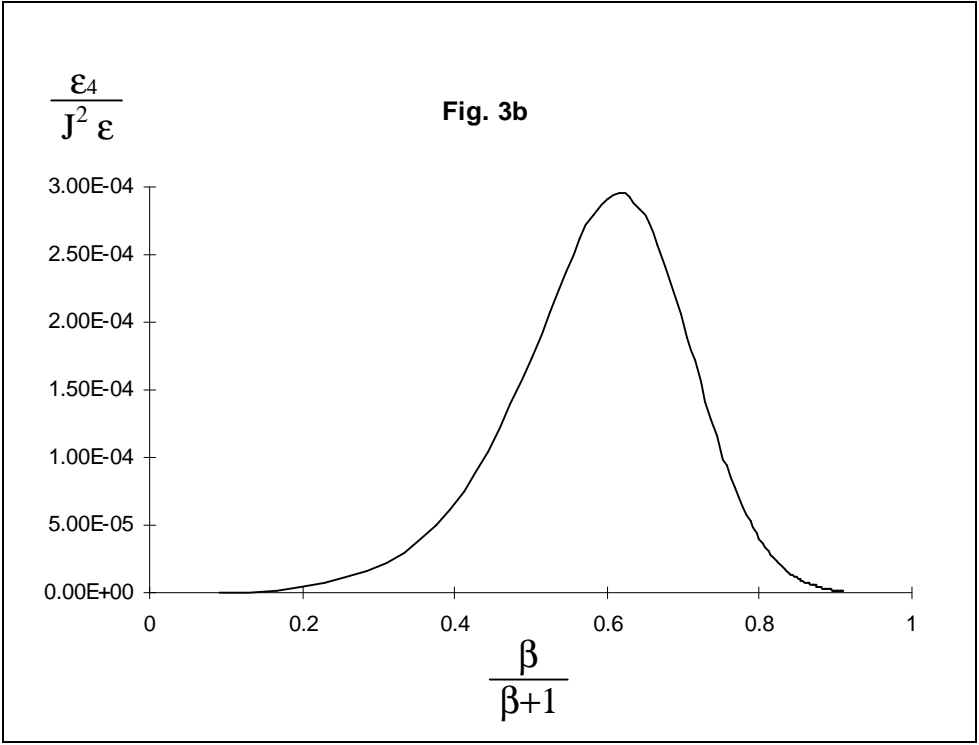
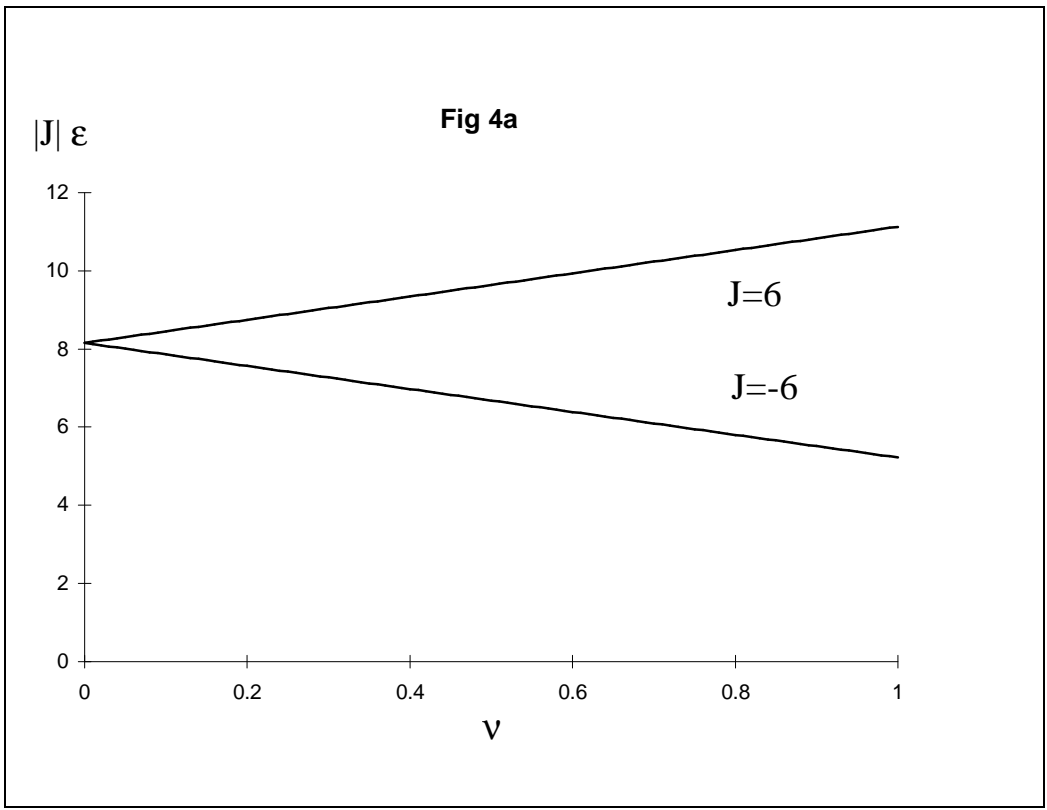
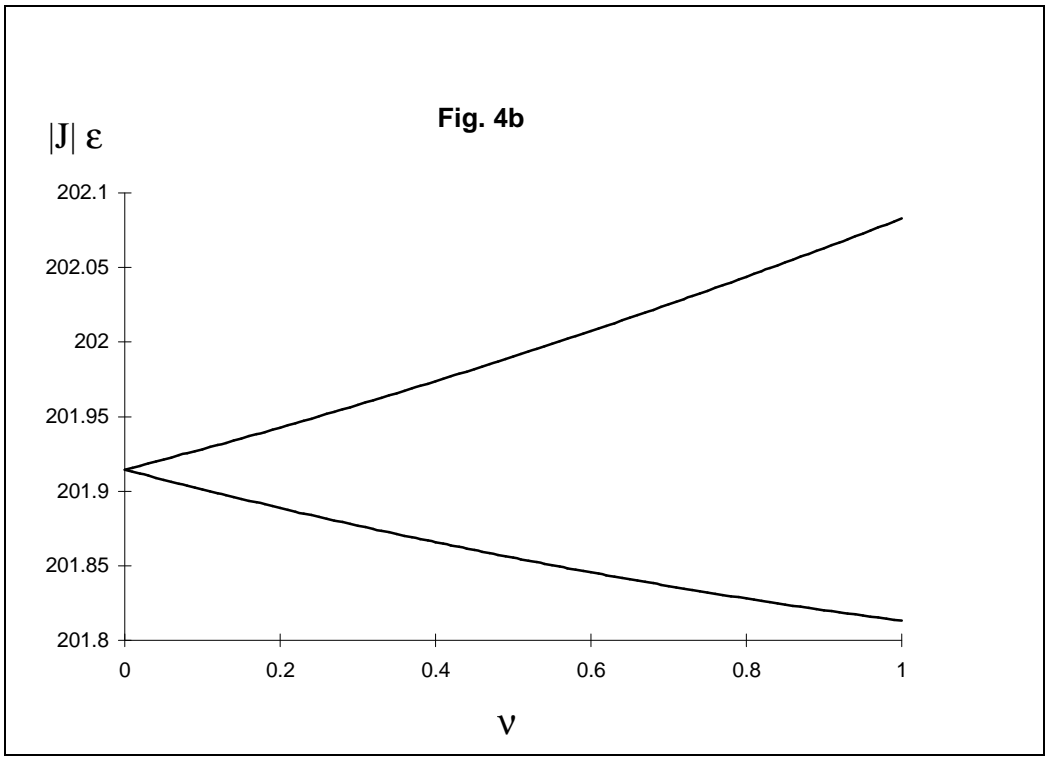
\end{document}